\documentclass[aps,twocolumn,10pt]{revtex4} 
\usepackage{amsmath}
\usepackage{graphicx}  
\usepackage{dcolumn}   
\usepackage{bm}        
\usepackage{amssymb}   
\usepackage{epstopdf}	 
\usepackage{placeins}
\usepackage{newfloat}
\usepackage{algorithm}
\usepackage{algpseudocode}

\begin{document}
\title{Honeycomb Lattices with Defects}
\date{\today}

\author{Meryl A. Spencer}
\affiliation{Department of Physics,\\University of Michigan, Ann Arbor MI 48104, USA}
\author{Robert M. Ziff}
\affiliation{Center for the Study of Complex Systems and Department of Chemical Engineering,\\University of Michigan, Ann Arbor MI 48109-2136, USA}

\begin{abstract}
In this paper we introduce a variant of the honeycomb lattice in which we create defects by randomly exchanging adjacent bonds, producing a random tiling with a distribution of polygon edges.  We study the percolation properties on these lattices as a function of the number of exchanged bonds using a novel computational method. We find the site and bond percolation thresholds are consistent with other three-coordinated lattices with the same standard deviation in the degree distribution of the dual; here we can produce a continuum of lattices with a range of standard deviations in the distribution. These lattices should be useful for modeling other properties of random systems as well as percolation. 
\end{abstract}

\maketitle

\section{Introduction} 
\begin{figure}[h]
\centering
\includegraphics[width=0.45\textwidth]{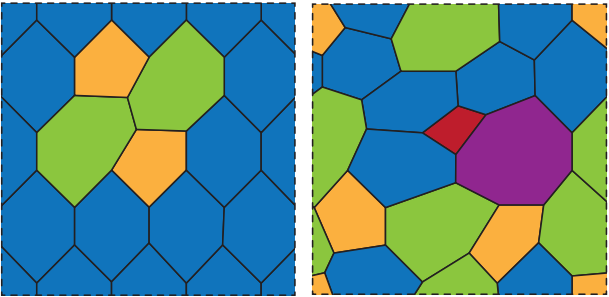}
\caption{Two examples of defective honeycomb (DHC) lattices. Each lattice has 16 tiles and periodic boundary conditions. Red tiles are rectangles, yellow tiles are pentagons, blue tiles are hexagons, green tiles are heptagons, and purple tiles are octagons. 
    				{\bf Left:} Example of a DHC lattice with standard deviation in degree of the dual $\sigma=0.5$. 
    				{\bf Right:} Example of a DHC lattice with $\sigma=1.658$.}
\label{fig:intro}
\end{figure}

Honeycomb lattices are extremely widespread, forming the structure of many natural and artificial objects. Atoms in graphene and other crystals are arranged in a honeycomb pattern, as are the bubbles that form in thin foams \cite{general03}. Consequently condensed matter and materials science researchers have a great interest in the properties of this lattice. Biologists study cells in epithelial sheets, which arrange themselves into honeycomb lattices \cite{BioReview}, not to mention the lattice's namesake, the beehive. The lattice is common in art and architecture too; you might even be sitting in a room with a honeycomb-tiled floor at this very moment!

The ideal honeycomb lattice used in art, architecture and math has been well studied (i.e., \cite{general01,general02,honeycomb_proof}), but naturally occurring systems are rarely perfect. Physical systems will almost always contain some defects which cause the systems to vary from the ideal lattice to some degree. In this paper we investigate how the severity of these defects affects the percolation properties of the lattice. We present a new lattice network which we call the ``Defective Honeycomb lattice" (DHC) formed by systematically swapping bonds in the honeycomb lattice, as shown in figure \ref{fig:intro}. We will describe how the basic properties of the lattice change as more bonds are flipped, as well as giving an algorithm for computationally creating such lattices. 

The percolation threshold describes the connectivity of a lattice. If sites or bonds are occupied with some probability $p$, the percolation threshold is the minimum value of $p$ at which there is a connected path from one side of the lattice to the other, in the limit of an infinite system \cite{StaufferAharony94}.  The percolation thresholds are known for many lattices, including the honeycomb lattice for bond percolation. In this paper we will compare the percolation thresholds of several known two-dimensional three-coordinated lattices with our DHC lattices, which are also three coordinated. We will show that there is a strong relationship between the percolation threshold and the variance in the number of faces in the polygons, equivalent to the degree of vertices on the lattice duals.

\section{Defective Honeycomb (DHC) Lattice}
Many naturally occurring biological systems, including cells in epithelial tissues, are approximate honeycomb lattices \cite{Yohans_pcp,dkl_lab}, but there is little understanding of the formal properties of these lattices. Understanding how defects in these lattices affects their properties will give us a more accurate picture of how naturally occurring honeycomb networks behave.

\subsection{Definition}
\begin{figure}
		\centering
    \includegraphics[width=0.4\textwidth]{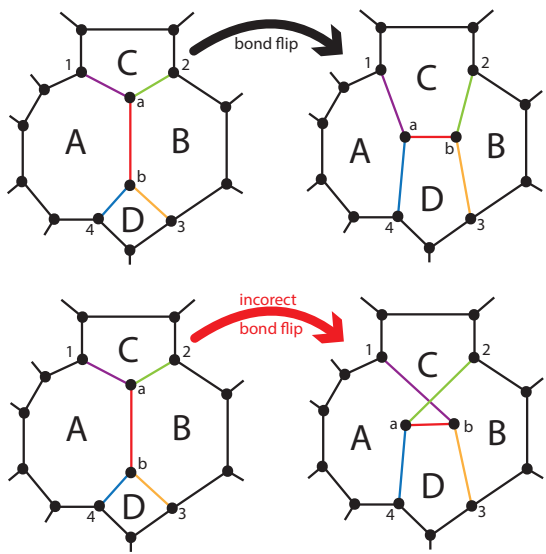}
    \caption{Pictorial representation of the flip of bond $a$--$b$ in the DHC lattice. The bonds between vertices $1$, $2$, $3$, $4$, $a$, and $b$, are rearranged such that tiles $C$ and $D$ become neighbors instead of $A$ and $B$. The top picture is a correct flip while the bottom picture shows and incorrect flip because it creates a non-planar lattice.  
    {\bf Top:} Correct flip. The bond $2$--$a$ becomes $2$--$b$ and the bond $4$--$b$ becomes $4$--$a$. The lattice remains planar and the bond $a$--$b$ separates two new tiles.
    {\bf Bottom:} Incorrect flip. There is no way to arrange sites $a$ and $b$ in the plane such that none of the bonds cross.}
    \label{fig:flip}
\end{figure}

The inspiration for the DHC lattice is the T1 topological process in foams \cite{general03}. In the T1 process one edge separating two bubbles shrinks down to a point and then regrows in a perpendicular direction. This has the effect of swapping which bubbles are neighbors, and the number of neighbors of the four bubbles involved. When viewed as a lattice this process is equivalent to rearranging the five bonds between six neighboring vertices connected in an H shape as shown in figure \ref{fig:flip}.

The DHC lattice is a honeycomb lattice in which a certain number of the lattice bonds have been flipped. The DHC lattice is defined by two parameters: $n$, the number of hexagonal tiles per row in the original $n\times n$ hexagonal lattice as arranged in figure \ref{fig:series} (top left), and $F$, the number of bonds which are flipped. To generate the lattice $F$ bonds are chosen uniformly at random; the same bond may be chosen more than once.  The order in which the bonds are flipped matters, as flipping bond $a$--$b$ and then $b$--$c$ is not equivalent to flipping bond $b$--$c$ and then $a$--$b$. 

In order to consistently describe lattices of different size we will specify $f=F/(3n^2)$, the fraction of edges flipped, instead of $F$, to characterize the number of flipped bonds in the lattice.

\subsection{Example}
\begin{figure*}
		\centering
    \includegraphics[width=0.9\textwidth]{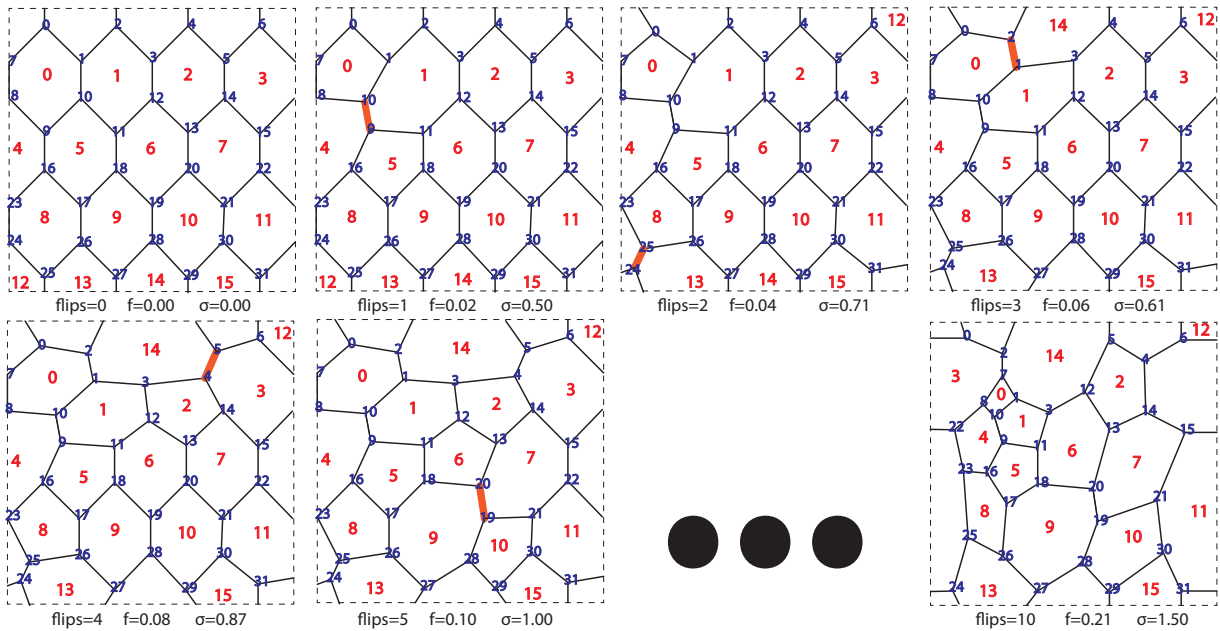}
    \caption{Pictures of the generation of an $n=4$, $f=10/48=0.21$ lattice. Ten distinct edges are flipped one at a time to generate the final lattice. Each panel shows the state of the lattice after each edge flip. The flipped edge is highlighted in red. The sites and tiles are labeled in blue and red respectively as described in section \ref{apx:init}. }
    \label{fig:series}
\end{figure*}

Figure \ref{fig:series} shows the generation of a $4\times4$ lattice with 10 flips, so $f=10/48\approx0.21$.   In the first panel none of the edges have been flipped and the system is the original honeycomb configuration. In the second panel the first bond chosen, bond $9$--$10$ shown in red, has been flipped. This has the effect of creating two pentagonal tiles and two heptagonal tiles in the otherwise unperturbed honeycomb lattice, increasing the standard deviation in the number of edges per tile, $\sigma$.   As the edges are flipped the lattice becomes more and more deformed and  $\sigma$ increases to $1.50$. The quantity $\sigma$, which is the same as the standard deviation in the coordination number of sites on the dual lattice, is important since it gives a quantitative measure of the extent of the difference between the new lattice and the original honeycomb.   It is defined as  
\begin{equation}
\sigma = \sqrt{\frac{1}{n^2}\sum_i (z_i-6)^2}
\label{standard deviation}
\end{equation}
where $z_i$ is the number of edges of the $i$-th polygon face, equal to the number of neighbors of the $i$-th vertex in the dual lattice.  The number 6 represents the mean value of $z_i$, which follows from Euler's formula, vertices$-$edges$+$faces $= 2$, assuming $n$ is large, and also from the fact that with each flip, the total number of bonds, tiles, and vertices is unchanged from the original honeycomb values. We also consider the variance $v=\sigma^2$.

\subsection{Properties}
There are several important properties of the DHC lattice. The lattice is two dimensional, so the percolation critical exponents of the system are known \cite{StaufferAharony94}. When a bond is flipped, four tiles are affected. Two of them (tiles $A$ and $B$ in figure \ref{fig:flip}) lose a bond and go from $n$-gons to $(n-1)$-gons, while  the other two (tiles $C$ and $D$ in figure \ref{fig:flip}) go from $n$-gons to $(n+1)$-gons. Therefore the average number of edges per tile (equivalently the average degree on the dual lattice) remains constant at 6.  As bonds are flipped the standard deviation in the number of edges per tile changes. The flipping of a bond changes the connection between sites, but each site continues to have exactly 3 neighbors.  Equivalently the dual lattice remains fully triangulated. We will use these 
 properties to define a class of planar lattices, the honeycomb variant class, where each lattice is two-dimensional, three-coordinated and has an average dual degree of six. We will compare the percolation thresholds of the DHC to the rest of the honeycomb variant class of lattices in section \ref{sec:pc}.

\section{Generation of Lattices}
\label{sec:comp}
In the remainder of this paper we will determine additional properties of the DHC lattices through computational simulation. The process of generating the lattices is mostly straightforward, however there are a few non-trivial components of the algorithm which we will address here. 

For physical reasons, we would like our lattice to remain planar.  Furthermore, for percolation, the critical exponents have different values in different dimensions, and may change if the lattice is non-planar. As we flip bonds we need a way to guarantee that our lattice stays planar. The specific coordinates of the sites are not relevant, but for an acceptable bond flip there must exist some arrangement of the sites on the plane for which none of the bonds cross. It turns out that the need for a planar representation of the lattice specifies a unique exchange of bonds for any flip trial. Figure \ref{fig:flip} shows the two topologically distinct ways to flip bond $a$--$b$. In both resulting lattices tiles $A$ and $B$ lose bond $a$--$b$  and tiles $C$ and $D$ gain bond $a$--$b$; however they differ in the reconnection of the bonds between sites $1$, $2$, $3$ and $4$, and the flipped sites $a$ and $b$. In the top picture bond $2$--$a$ is broken and replaced by bond $2$--$b$ and bond $4$--$b$ is broken and replaced by bond $4$--$a$. In the bottom picture bond $1$--$a$ is replaced by bond $1$--$b$ and bond $4$--$b$ is replaced by $4$--$a$. The first set of reconnections are correct because the lattice has a planar representation which is shown. In the bottom case there is no planar representation on the lattice---no matter where sites $a$ and $b$ are located there will be some bonds that cross each other---therefore we need to be careful when reconnecting the bonds after a flip to preserve the planar nature of the lattice.  

Lattices are often stored on computers as simple neighbor lists, however here a neighbor list would not be able to determine which of the two distinct ways to flip bond $a$--$b$ would result in a planar representation. In each case sites $a$ and $b$ exchange one neighbor with each other, but without knowledge of which neighbors belong to which tiles it is impossible to determine if sites $2$ and $4$ should be exchanged or sites $1$ and $3$. In order to solve this problem we stored our lattices as a list of all of the sites that make up each tile. The sites were stored in clockwise order to further simplify the flipping algorithm. For example ${\tt tilelist}[3]=[1,5,2,18,7]$ denotes that the tile with ID of 3 is a pentagon made of the sites 1, 2, 5, 7, 18, and bonds $1$--$5$, $5$--$2$, $2$--$18$, $18$--$7$ and $7$--$5$.  Storing the lattice in this way makes the algorithm given in algorithm \ref{alg:flip} for flipping bond $a$--$b$ simple. It is also easy to translate into the standard neighbor list as the three neighbors of site 2 are all of the IDs immediately following 2 in the entire tile list. 

In order to use algorithm \ref{alg:flip} to flip bonds in the lattice we must start with the lattice as lists of clockwise sites in every cell. We used the labeling method shown in figure \ref{fig:series} to determine tile and site IDs. Tiles are initially labeled 0 to $n^2-1$ in order by row and then column. The top site of each hexagon initially has an ID of ID$_\mathrm{site}=2 ($ID$_\mathrm{tile})$ and the next clockwise site has ID$_\mathrm{site}=2($ID$_\mathrm{tile})+1$. This produces a label for every site and tile in the lattice. The lattice is then initialized according to the algorithm given in appendix \ref{apx:init}.  

\begin{algorithm}
\caption{Flip Bond a-b}
  \label{alg:flip}
\begin{algorithmic}[1]
\State{A=tile with site list ...$a$,$b$,... } \Comment{Find the four tiles involved in the flip}
\State{B=tile with site list ...$b$,$a$,... }
\State{C=tile with site list ...$a$,... }
\State{D=tile with site list ...$b$,...}
\State{A $\rightarrow$ delete($b$) } \Comment{Change the bonds }
\State{B $\rightarrow$ delete($a$) }
\State{C $\rightarrow$ insert($b$) before $a$ }
\State{D $\rightarrow$ insert($a$) before $b$}
\end{algorithmic}
\end{algorithm}

\section{Relationship between flips and defects}
\begin{figure}
		\centering
    \includegraphics[width=0.45\textwidth]{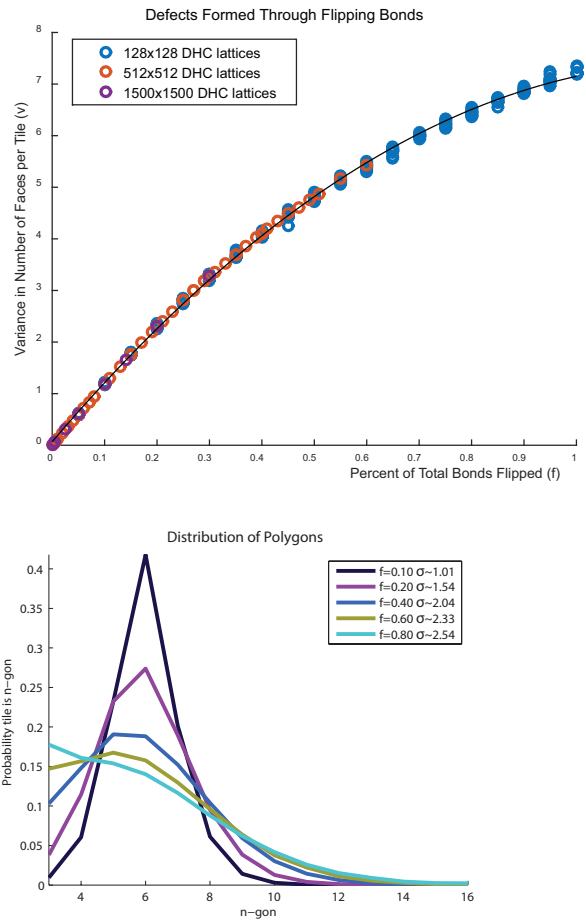}
    \caption{Relationship between the variance in the number of edges per tile and the fraction of bonds flipped $f$. {\bf Top:} Blue circles represent the variance found in 200 independently generated $128\times128$ tile lattices. Orange circles represent the variance found in 50 independently generated $512 \times 512$ tile lattices. Purple circles represent the 8 $1500\times1500$ tile lattices used to find the percolation thresholds in section \ref{sec:pc}. The data follows the function $v=-4.835f^2+11.92f+0.0681$. {\bf Bottom:} Average distribution of polygons over ten independent $128\times128$ lattices at different flipping fractions. As more bonds are flipped a greater percentage of the tiles are triangles and many-sided polygons.}
    \label{fig:flips}
\end{figure}

The DHC lattice is characterized by the number of bonds flipped, and we would like to know how the lattice transforms in the process. We flip the bonds randomly, so we will look at the average behavior over many lattices. To quantify the `defectiveness' of an individual lattice we calculate the $\sigma$ of the degree of the dual lattice (\ref{standard deviation}). We created a set of independent DHC lattices of three different sizes, $n = 128, 512$ and $1500$ and compared the number of flips the lattice had undergone to the standard deviation of defects $\sigma$. The results are shown in figure \ref{fig:flips}. There is a very strong relationship between the extent of defects as characterized by the variance $v$ and the normalized number of bonds flipped, described approximately by $v=-4.835f^2+11.92f+0.0681$.  

We also want to characterize the qualitative behavior of the system as the number of flipped bonds increases. We measured the probability of a tile having $n$ sides as a function of $f$.  At small $f$ the tiles are almost all hexagons, with a few pentagons and heptagons. As more of the bonds are flipped the percent of hexagonal tiles in the lattice decreases as shown in figure \ref{fig:flips} (right). The long-term behavior of the system is to create numerous triangular tiles and a few many-sided polygons. The reason for this long-time behavior is that every time a bond is flipped two tiles lose an edge and two tiles gain an edge, so as more bonds are flipped the number of bonds per tile is forced away from the mean of six. 

There is an ambiguity in the definition of the lattice when $f$ gets large. If the next bond to be flipped is part of a triangle it cannot be flipped or a tile with only two sites would be created and the lattice would become a multi-graph (where two sites are connected by more than one edge) which is not the behavior we were looking for.  Bonds that are part of triangular tiles should not be flipped, but it is up to us to define whether choosing an unflippable bond counts as a flip or not. We in fact chose to count those bonds towards the flip trial count of the lattice. 

We were not able to fully determine the long-time behavior of the infinite DHC system due to a specific finite-size effect. We used finite latices with periodic boundary conditions to approximate the infinite system. As the bonds in the system are flipped the tiles which neighbor each other change, and eventually one tile will gain enough bonds that it will wrap around the finite system and neighbor itself, dividing the lattice into two. When this occurred we stopped the simulation as the lattice was no longer properly defined.

\section{Determination of $p_c$}
\label{sec:pc}

\begin{table*}
  \centering
  \caption{Honeycomb variant (three-coordinated) lattices}
	\begin{tabular}{ | l | l | c | c | c | }
    \hline
    Lattice & Common Name & $\sigma$ & \phantom{spacess} $p_c^\mathrm{site}$ \phantom{spacess}& \phantom{spacess} $p_c^\mathrm{bond}$ \phantom{spacess} \\ \hline \hline
    \phantom{m}$(6^3)$ \phantom{m}& \phantom{m}honeycomb	& 0.000	& 	0.697040 \cite{general01}	& 	0.652704 \cite{SykesEssam64}\\ \hline
    \phantom{m}- & \phantom{m}Poisson-Voronoi \phantom{m}	& 	1.314 \cite{Hilhorst07} &	0.71410 \cite{thresh_pv} & 0.66693 \cite{thresh_pv}\\	\hline
    \phantom{m}$(4,8^2)$ 	\phantom{m}& \phantom{m}bathroom tile 	& 2.061	&	0.729723 \cite{Jacobsen14} &0.676803 \cite{Jacobsen14} \\ \hline
    \phantom{m}$(4,6,12)$ \phantom{m}	& \phantom{m}cross 	& 	2.828 	&	0.747801 \cite{Jacobsen14}&0.693731 \cite{Jacobsen14}\\ \hline
    \phantom{m}$\frac{3}{4}(3,9^2)+\frac{1}{4}(9^3)$ 	\phantom{m}& \phantom{m}martini & 3.000 &0.764826 \cite{Scullard06}& 0.707107\cite{thresh_martini}\\ \hline
    \phantom{m}$(3, 12^2)$ \phantom{m}& \phantom{m}three-twelve	& 4.235	&	0.807901 \cite{SudingZiff99}	&	0.740421 \cite{Jacobsen14}\\ 
    \hline
\end{tabular}
\label{table:names}
\end{table*}

\begin{table}
  \centering
  \caption{Site and bond thresholds for the DHC lattices from simulations on 1500 x 1500 systems}
	\begin{tabular}{|c | c | c|}
    \hline
    \phantom{m} $\sigma$ \phantom{m} &  \phantom{spacess} $p_c^\mathrm{site}$ \phantom{spacess}& \phantom{spacesp} $p_c^\mathrm{bond}$ \phantom{spacesp}\\ \hline \hline
			0.24 	&	$0.69770 \pm  0.00025$ 	& $0.65320 \pm  0.00025$\\ \hline
			0.55	&	$0.69995 \pm  0.00015$	&	$0.65497\pm  0.00025$\\ \hline
			0.77	&	$0.70200 \pm  0.00025	$&	$0.65715\pm  0.00025$\\ \hline
			1.09	&	$0.7078 	\pm  0.00004$	&	$0.66130\pm  0.00025$\\ \hline
			1.28	&	$0.71075	\pm  0.00025$	&	$0.66440\pm  0.00025$\\ \hline
			1.52	&	$0.71659	\pm  0.00004$	&	$0.66860\pm  0.00025$\\ \hline
			1.81	&	$0.72380 \pm  0.00004$	&	 $0.67470\pm  0.00025$\\ \hline
			2.10	&	$0.73147 \pm  0.00012$ 	&	 $0.68140\pm  0.00025$\\  
    \hline
\end{tabular}
\label{table:pcdhc}
\end{table}

\begin{figure*}
		\centering
    \includegraphics[width=0.7\textwidth]{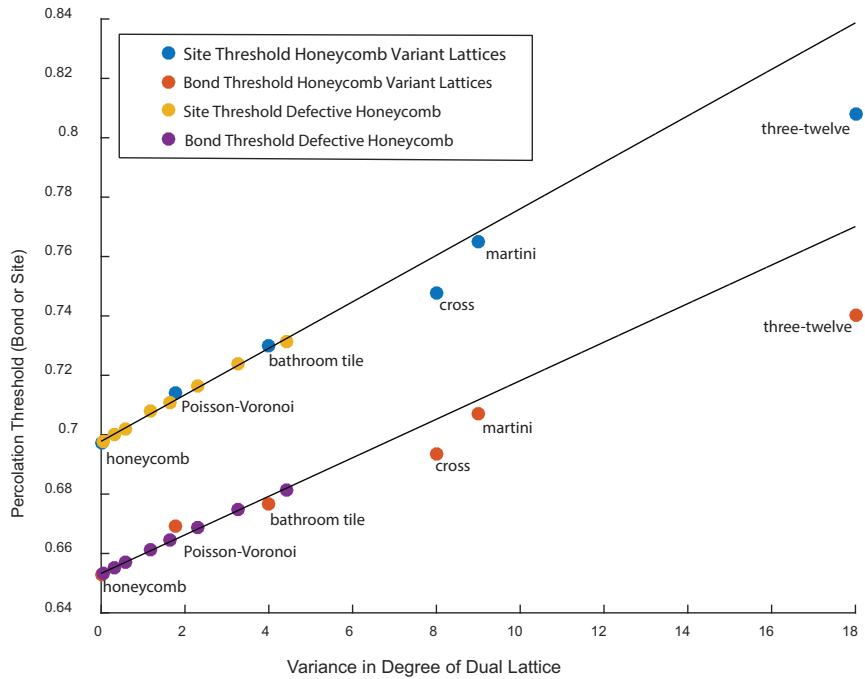}
    \caption{Graph of the relationship between the percolation threshold and dual-lattice variance $v=\sigma^2$. The bond (lower) and site (upper) thresholds for the DHC lattice and the previously studied honeycomb variant lattices listed in table \ref{table:names} are shown. The linear regression for the bond thresholds is $p_c=0.00649v+0.6533$ and for the site threshold is $p_c=0.00783v+0.6978$.}
    \label{fig:pcdata}
\end{figure*}

We found the percolation thresholds of the DHC computationally using a method described fully in Appendix \ref{apx:pc}. At the percolation threshold  the ratio
\begin{equation}
y(s,p)=\frac{s P_{\geq s}}{\langle{s' {\scriptstyle \leq } s} \rangle}
\end{equation}
reaches a constant value of $\delta-1$, for large $s$, where $s$ is the size of a cluster, ${\langle{s' {\scriptstyle \leq } s} \rangle}$ is the expectation value of $s$ for clusters up to size $s$, $P_{\geq s}$ is the probability that a vertex is in a cluster of at least size $s$, and $\delta = 1/(\tau-2) = 91/5$ (in two dimensions) is a critical exponent of percolation. By graphing $y(s,p)$ for different values of the occupation probability $p$ and finding when it reached $17.2$ for large $s$, we were able to determine $p_c$ with high precision. 

To find the thresholds, we carried out 100,000 simulations on lattices of size $n = 1500$, for values of $\sigma$ equal to 0.00, 0.24, 0.55, 0.77, 1.09, 1.28, 1.52, 1.81, and 2.10 for both site and bond percolation.  The details are shown in figure \ref{fig:pcdata}.  

The final results are shown in Figure \ref{fig:pcdata}, which gives the relationship between the $p_c$ and $\sigma$ of the lattices.  We also include in that plot the known thresholds of the honeycomb variant lattices (other lattices with coordination number 3).   As $\sigma$ increases both the site and bond percolation thresholds increase in a fairly predictable manner.  The honeycomb variant lattices we used, which were all the three-coordinated ones where thresholds are known, are summarized in table \ref{table:names}.

\section{Discussion}
We introduced the concept of the DHC lattice as a way to better model real-world lattices which are not perfect honeycombs. The lattices are generated probabilistically by choosing edges uniformly at random from a honeycomb lattice to 'flip'.  This process creates a new lattice with increasing number of defects as the bonds are flipped. We can quantify the severity of the defects by measuring $\sigma$, the standard deviation, or $v=\sigma^2$, the variance in the number of edges per tile. The variance is linearly proportional to the fraction of edges flipped $f$ for small $f$. As more edges are flipped the distribution of polygons in the lattice changes such that most tiles are triangles or have a large number of edges. 

There is a series of well-studied three-coordinated lattices which have many of the same basic properties of the DHC lattices. We compared the percolation threshold of these lattices and found that the percolation threshold for all of the lattices increases linearly in $v$. This means that the variance is a useful quantity to use to determine the percolation threshold of this class of lattice.  The increase of thresholds as $v$ increases means that ideal honeycomb lattice is more robust to connecting paths through the network than its defective counterparts. Patterning in biological systems takes energy and so we would expect that patterns like the honeycomb should provide some advantage over disordered lattices in order to be evolutionarily favorable. The fact that the honeycomb pattern is more robust against site or bond failures than similar networks with defects may be part of the reason it is so common in the natural world. 

\FloatBarrier
\bibliographystyle{ieeetr}
\bibliography{master}

\begin{thebibliography}{10}

\bibitem{general03}
D.~Weaire and S.~Hutzler, {\em {The Physics of Foams}}.
\newblock {Clarendon Press}, {1999}.

\bibitem{BioReview}
A.~G. Fletcher, M.~Osterfield, R.~E. Baker, and S.~Y. Shvartsman, ``Vertex
  models of epithelial morphogenesis,'' {\em Biophysical J.}, vol.~106, no.~11,
  pp.~2291 -- 2304, 2014.

\bibitem{general01}
J.~L. Jacobsen, ``{High-precision percolation thresholds and Potts-model
  critical manifolds from graph polynomials},'' {\em {J. Phys. A: Math. Th.}},
  vol.~{47}, {APR 4} {2014}.

\bibitem{general02}
T.~Hales, ``{The honeycomb conjecture},'' {\em {Discrete \& Computational
  Geometry}}, vol.~{25}, pp.~{1--22}, {JAN} {2001}.

\bibitem{honeycomb_proof}
H.~Duminil-Copin and S.~Smirnov, ``{The connective constant of the honeycomb
  lattice equals root 2+root 2},'' {\em {Annals of Mathematics}}, vol.~{175},
  pp.~{1653--1665}, {MAY} {2012}.

\bibitem{StaufferAharony94}
D.~Stauffer and A.~Aharony, {\em {Introduction to Percolation Theory, 2nd ed.}}
\newblock CRC Press, 1994.

\bibitem{Yohans_pcp}
P.-L. Bardet, B.~Guirao, C.~Paoletti, F.~Serman, V.~Leopold, F.~Bosveld,
  Y.~Goya, V.~Mirouse, F.~Graner, and Y.~Bellaiche, ``{PTEN} controls junction
  lengthening and stability during cell rearrangement in epithelial tissue,''
  {\em {Developmental Cell}}, vol.~{25}, pp.~{534--546}, {JUN 10} {2013}.

\bibitem{dkl_lab}
P.~A. Raymond, S.~M. Colvin, Z.~Jabeen, M.~Nagashima, L.~K. Barthel,
  J.~Hadidjojo, L.~Popova, V.~R. Pejaver, and D.~K. Lubensky, ``Patterning the
  cone mosaic array in zebrafish retina requires specification of
  ultraviolet-sensitive cones,'' {\em {PLOS ONE}}, vol.~{9}, {JAN 21} {2014}.

\bibitem{SykesEssam64}
M.~F. Sykes and J.~W. Essam, ``Exact critical percolation probabilities for
  site and bond problems in two dimensions,'' {\em J. Math. Phys.}, vol.~5,
  no.~8, 1964.

\bibitem{Hilhorst07}
H.~J. Hilhorst, ``New {M}onte {C}arlo method for planar {P}oisson-{V}oronoi
  cells,'' {\em J. Phys. A: Math. Th.}, vol.~40, no.~11, p.~2615, 2007.

\bibitem{thresh_pv}
A.~M. Becker and R.~M. Ziff, ``Percolation thresholds on two-dimensional
  {V}oronoi networks and {D}elaunay triangulations,'' {\em Phys. Rev. E},
  vol.~80, p.~041101, Oct 2009.

\bibitem{Jacobsen14}
J.~L. Jacobsen, ``{High-precision percolation thresholds and Potts-model
  critical manifolds from graph polynomials},'' {\em J. Phys. A: Math. Th.},
  vol.~47, p.~135001, Apr. 2014.

\bibitem{Scullard06}
C.~R. Scullard, ``Exact site percolation thresholds using a site-to-bond
  transformation and the star-triangle transformation,'' {\em Phys. Rev. E},
  vol.~73, p.~016107, Jan 2006.

\bibitem{thresh_martini}
R.~M. Ziff, ``Generalized cell--dual-cell transformation and exact thresholds
  for percolation,'' {\em Phys. Rev. E}, vol.~73, p.~016134, Jan 2006.

\bibitem{SudingZiff99}
P.~N. Suding and R.~M. Ziff, ``Site percolation thresholds for {A}rchimedean
  lattices,'' {\em Phys. Rev. E}, vol.~60, pp.~275--283, Jul 1999.

\bibitem{Leath76}
P.~L. Leath, ``Cluster size and boundary distrabution near percolation
  threshold,'' {\em Phys. Rev. B}, 1976.

\bibitem{NewmanZiff00}
M.~E.~J. Newman and R.~M. Ziff, ``Efficient {M}onte {C}arlo algorithm and
  high-precision results for percolation,'' {\em Phys. Rev. Lett.}, vol.~85,
  pp.~4104--4107, Nov 2000.

\bibitem{NewmanZiff01}
M.~E.~J. Newman and R.~M. Ziff, ``Fast {M}onte {C}arlo algorithm for site or
  bond percolation,'' {\em Phys. Rev. E}, vol.~64, p.~016706, Jun 2001.

\end{thebibliography}

\onecolumngrid
\appendix

\section{Initializing the Lattice}
\label{apx:init}

In order to use algorithm \ref{alg:flip} the lattice must be stored as a list of clockwise-ordered vertices around each cell. There are many different solutions which give unique labels to every vertex, and working out the proper boundary conditions and determining the vertices in each tile can be time consuming. We have reproduced our algorithm here for interested parties wishing to produce their own DHC lattices. We used the labeling method shown in figure \ref{fig:series} to determine tile and site IDs. Tiles are labeled 0 to $n^2-1$ in order by row and then column. The top site of each hexagon has id ID$_{site}=2 ($ID$_{tile})$ and the next clockwise site has ID$_{site}=2 ($ID$_{tile})+1$. The following pseudocode produces a two-dimensional array of $n\times n$ hexagons. The $i$th row of the array gives the clockwise ID's of vertices of tile $i$ where the first entry is the top vertex of the tile. 

\begin{algorithm}
\caption{Initialize Lattice}
  \label{alg:init}
\begin{algorithmic}[1]
\For{z=0 to $n^2-1$}
	\State{$\alpha= \frac{z - (z \pmod n)}{n} \pmod 2$}	
	\State{$z'=2(z+n \pmod n)$} 
	\State{tiles$[z][0]=2z$} 
	\State{tiles$[z][1]=2z+1$} 
	\If{$\alpha =1$ }
		\State{tiles$[z][2]=z'$}
		\If{ $z \pmod n =0$}
			\State{tiles$[z][3]=2(z+2n-1 \pmod n^2)+1$}
			\State{tiles$[z][4]=2(z+2n-1 \pmod n^2)$}
		\Else
			\State{tiles$[z][3]=z'-1$}
			\State{tiles$[z][4]=z'-2$}
		\EndIf
	\EndIf
	\If{$z \pmod n = n-1$}
		\State{tiles$[z][2]=2z+2$}
		\State{tiles$[z][2]=z'+2$}
	\Else
		\State{tiles$[z][2]=z'+1$}
		\State{tiles$[z][2]=z'$}
	\EndIf
	\If{ $z \pmod n \neq 0$}
		\State{tiles$[z][5]=2z-1$}
	\Else
		\State{tiles$[z][5]=2(z+n-1)+1$}
	\EndIf
\EndFor
\end{algorithmic}
\end{algorithm}
\FloatBarrier
\section{Determining PC}
\label{apx:pc}
\begin{figure*}
		\centering
    \includegraphics[width=0.95\textwidth]{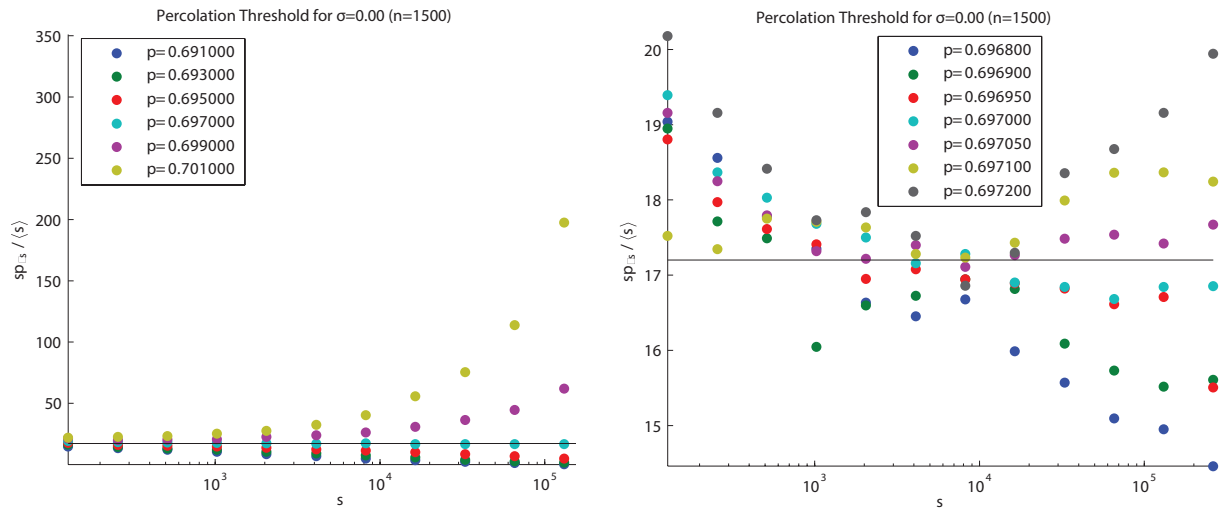}
    \caption{Example of finding the percolation threshold using our method. We grew 10000 clusters on a honeycomb lattice at various occupation probabilities $p$. Statistics on the average $P_{\geq s}$ and $\langle s' \leq s \rangle$ were gathered and the ratio $y= s P_{\geq s} / {\langle s' \leq s \rangle}$ was plotted versus the cluster size $s$. At the percolation threshold the resulting curve should be flat. On the left we get the approximate value of $p_c$ for site percolation on the honeycomb lattice. On the right we narrow in on the exact value. Using this method we determined that the site percolation threshold was $0.69700 \pm 0.00005$ which agrees with the known value of 0.697040 \cite{general01}.}
    \label{fig:expc}
\end{figure*}

In section \ref{sec:pc} we presented data showing the relationship between the percolation threshold of some DHC lattices with the honeycomb variant lattices. The site and bond percolation thresholds for the honeycomb variant lattices are known (see figure \ref{table:names}). There are many established ways to estimate the percolation threshold of a lattice using simulations \cite{StaufferAharony94,Leath76}, however we found that a novel method based on a cluster growing algorithm was more practical for our problem. In this section we will describe the theory behind our method and outline the computational model we used. 

The theory behind the method is straightforward. We found a ratio of properties which we will call $y$ (defined below) that is a constant function of $s$ (for large $s$) at the percolation threshold. We can then use a normal cluster growing algorithm to determine this ratio at different occupation probabilities, and narrow in on the value of $p_c$. 

The following is a short theoretical derivation proving that our ratio $y$ should be constant for $s$ at the percolation threshold. Let $n_s$ be the number of clusters of size $s$, and $P_s$ be the probability that any site is in a cluster of size $s$. It is well known \cite{StaufferAharony94} that at $p_c$ the number of clusters of size $s$, is $As^{-\tau}$, where $A$ is some constant and $\tau$ is the Fisher scaling exponent. From this relationship it follows that the probability that a site on the lattice belongs to a cluster of size $s$ is given by
\begin{equation}
P_{s}=sn_s=A s^{1-\tau}.
\end{equation} 
\label{eq:ps}
which implies that the probability a site is in a cluster of at least size $s$ is given by 
\begin{equation}
P_{\geq s}=\int_{s}^{\infty}{P_{s'} ds'}=\int_{s}^{\infty}A({s'})^{1-\tau}ds',
\end{equation} 
\label{eq:pgeqs}
at $p_c$. The value of $\tau$ is between $2$ and $2.5$ in every dimension so the integral in (\ref{eq:pgeqs}2) converges and  evaluates to
\begin{equation}
P_{\geq s}=\frac{As^{2-\tau}}{\tau-2}.
\end{equation} 
We want our ratio of lattice properties $y$ to be constant at $p_c$, so we incorporated the average value of the size of all clusters up to size $s$:
\begin{equation}
\langle{s' {\scriptstyle \leq } s} \rangle=\int_{1}^{s}s'P_{s'}=\int_{1}^{s}A(s')^{2-\tau}ds'=\frac{As^{3-\tau}-1}{3-\tau}.
\label{eq:average}
\end{equation} 
It follows that the ratio $y$, defined as  
\begin{equation}
y=\frac{sP_{\geq s}}{\langle{s' {\scriptstyle \leq } s} \rangle}=\frac{3-\tau}{\tau-2}=\frac{86}{5}.
\end{equation} 
is constant for large $s$.

In order to use the ratio $y$ to find the value of $p_c$ we ran a standard epidemic cluster growing algorithm with occupation probability $p$ and recorded the number of sites in each cluster until appropriate statistics had been collected.  Clusters were grown until a preset maximum cutoff $s_\mathrm{max}$ was hit.  Data were binned by $\log_2{s}$, that is, bins for $s =  1,$ 2-3, 4-7, 8-15, $ \dots$ and those that hit $s_\mathrm{max}$, with bins both for the number in each bin, and the total $s$ of all the clusters in each bin.  Summing these bins respectively above $s$ and below $s$ allowed us to calculate $y$ for $s = 2, 4, 8, \ldots$, $s_\mathrm{max}$. 

When the occupation threshold is above $p_c$ the curve for $f$ is above $17.2$, and when the occupation threshold is below $p_c$ the curve is below $17.2$. Therefore, we can narrow in on the percolation threshold by testing various values of $p$ and graphing them as shown in figure \ref{fig:expc}. First we scan through a range of $p$ values using a large step size and short simulations to determine the approximate $p_c$. We then run extensive simulations on just a few values of $p$ near the threshold to get an accurate value of $p_c$.  This method is easy to program and efficient when one is trying to find just the threshold, and not the behavior of the system for all values of $p$, in which case other methods \cite{NewmanZiff00,NewmanZiff01} are more efficacious.

For this method to work well, the lattice has to be big enough so that the maximum cluster size $s_\mathrm{max}$ could be reached before the boundaries are hit.  Then there are no finite-size effects due to the boundaries.  However, there are still finite-size (lattice) effects for small clusters; these are generally unimportant for $s > 1000$.

\end{document}